\documentclass[twocolumn, 10pt]{article}

\usepackage[left=1.25cm, right=1.25cm, top=2cm, bottom=2cm]{geometry}
\usepackage{hyperref}
\usepackage[T1]{fontenc}
\usepackage{times}
\usepackage{color}
\usepackage{amsmath, amssymb, amsthm}
\usepackage[]{algorithm2e}
\usepackage{graphicx}
\usepackage{subfigure}
\usepackage{booktabs}

\newcommand{\Z}{\mathbf{Z}}

\newtheorem{newdef}{Definition}
\newtheorem{proposition}{Proposition}

\begin{document}

\title{Randomized Experimental Design via Geographic Clustering}

\author{David Rolnick\footnote{University of Pennsylvania, drolnick@seas.upenn.edu},\, Kevin Aydin\footnote{Google, \{kaydin, jeanpa, kamali, mirrokni, amir\}@google.com},\, Jean Pouget-Abadie\footnotemark[2],\\
Shahab Kamali\footnotemark[2],\, Vahab Mirrokni\footnotemark[2],\, Amir Najmi\footnotemark[2]}
\date{}

\maketitle

\begin{abstract}
Web-based services often run randomized experiments to improve their products. A popular way to run these experiments is to use geographical regions as units of experimentation, since this does not require tracking of individual users or browser cookies. Since users may issue queries from multiple geographical locations, geo-regions cannot be considered independent and interference may be present in the experiment. In this paper, we study this problem, and first present GeoCUTS, a novel algorithm that forms geographical clusters to minimize interference while preserving balance in cluster size. We use a random sample of anonymized traffic from Google Search to form a graph representing user movements, then construct a geographically coherent clustering of the graph. Our main technical contribution is a statistical framework to measure the effectiveness of clusterings. Furthermore, we perform empirical evaluations showing that the performance of GeoCUTS is comparable to hand-crafted geo-regions with respect to both novel and existing metrics.
\end{abstract}

\section{Introduction}
\label{sec:intro}

Large-scale online services routinely conduct live experiments to improve their products. As described in \cite{diane-overlapping}, browser cookies are the standard unit of analysis for experiments run by many web services. Typically, cookies are randomly selected into disjoint treatment and control groups and then subjected to different treatments (e.g.~background color). The statistical and practical significance of metric differences between the two groups of cookies are important factors in the decision whether or not to launch the experimental treatment to all users. While this cookie-based approach has been used extensively in the industry, it has some important limitations when measuring long-term effects on users. As detailed in \cite{henning}, the fundamental problem is that \emph{``[cookies]...are a poor proxy for users...Users can clear their cookies whenever they want, and they frequently use multiple devices and multiple browsers.''} The authors further observe that \emph{``Using a signed-in id may seem to mitigate this issue, but users can have multiple sign-ins and many queries are made while signed-out.''} Thus the effect measured by a cookie experiment is diluted by the fact that a user may be in the treatment group on one device or browser and in the control group in another.

To fix this problem, experiments are often run on geographical regions \cite{vaver}. This is a form of cluster-based randomization~\cite{imbens2015causal}. Assuming a user remains within one such region, the user will receive consistent experimental treatment regardless of which device or browser she uses. Geo-partitions are also somewhat well-suited to mitigate secondary effects from interactions between the treatment and control groups. When experiments are run with large, noticeable changes (e.g.~a new feature or significant redesign of an existing feature), it is often the case that users influence each other by word of mouth. Such influence tends to be biased towards geographically local interaction~\cite{backstrom2010find}. Geo-regions have typically been hand-designed with great effort. Moreover their design has typically ignored the actual movements of web users, which means an increased number of users move between regions, causing \emph{interference} between the treatment and control arms of the experiment, which violates the Stable Unit Treatment Value Assumption (SUTVA)~\cite{imbens2015causal} on which standard causal inference analyses rely.

In this paper, we aim to design a framework for running randomized experiments using geographic clustering. To this end, we first present a distributed algorithm, Geographic Clustering Using Travel Statistics (GeoCUTS) designed to mitigate interference whilst ensuring experimental power. We perform a comprehensive evaluation on massive quantities of Google Search data to study the impact of different design choices on GeoCUTS and compare its performance against alternatives, for both novel and existing metrics. To do so, we describe a statistical framework for evaluating the quality of the clusters, and a new metric---the Q-metric---for this purpose. This statistical framework and the new metric are of independent interest as they present a novel way to measure effectiveness of clustering for such experimental design problems. In particular, we formally show the relationship between minimizing the cut via a balanced partitioning and optimizing the Q-metric, which, in turn, we show represents the quality of experimental design. Finally, the results of our empirical study suggest that the performance of GeoCUTS is equal to or surpasses that of hand-designed regions.

\section{Prior results}
Many authors~\cite{aronow2013estimating, manski2013identification, saveski2017detecting} have considered the problem of causal inference in a network-structured domain. Experimental treatment is imposed upon some nodes and interference occurs between nodes that are sufficiently close in the network.  For example, in a social network, an individual's response may be influenced both by their own treatment and by that of their friends or friends of friends~\cite{athey2018exact,basse2015optimal, walker2014design, middleton2011unbiased}.  Ugander et al.~\cite{ugander2013graph} and Gui et al.~\cite{gui2015network} design low-variance estimators for these cases, relying upon clusters within the network. Eckles et al.~\cite{eckles2014design} demonstrate the effectiveness of \emph{network bucket testing}, randomly assigning treatments to different clusters, which gives a natural low-variance estimator. This approach is analyzed further by Backstrom and Kleinberg \cite{backstrom2011network} and by Katzir et al.~\cite{katzir2012framework}, who consider making clusters using weighted random walks on the network. Few of these prior works however consider the bipartite setting that we study here. Our main contribution to this literature is the introduction of the Q-metric and its justification as a suitable quality metric for causal experiments on bipartite graphs.

Furthermore, our paper builds upon existing literature for finding balanced clusters efficiently within a large network and evaluating the resulting clustering. In {\em balanced partitioning}, our goal is to find a set of clusters of almost equal size and to minimize the total weight of edges that cross clusters (i.e., minimize the cut).  This is an NP-hard problem that is computationally hard even for medium-sized graphs~\cite{AR06} as it captures the problem of graph bisection~\cite{gareyjohnson}. No constant approximation algorithm is known. Logarithmic time approximation algorithms are based on solving linear programming and semi-definite programming relaxations for these problems. Such relaxations are hard for graphs with thousands of nodes, yet we must consider still larger graphs, which require effective heuristics that can be implemented in a distributed manner.

While the topic of large-scale balanced graph partitioning has attracted significant attention in the literature~\cite{karypis1998multilevelk, DGRW12,DGRW11,tsourakakis2014fennel,ugander2013balanced,stanton2012streaming, nishimura2013restreaming, nelson2016economic}, many prior authors have studied large-scale but non-distributed solutions to this problem. The need for distributed algorithms has been observed by several practical and theoretical research papers~\cite{aydin2015distributed,tsourakakis2014fennel,ugander2013balanced}. Zhu and Ghahramani \cite{zhu2002learning} introduced the approach of label propagation, which was generalized to balanced clusters in the work of Ugander and Backstrom \cite{ugander2013balanced}. More recently, Aydin et al.~\cite{aydin2015distributed} achieve a scalable clustering in large networks by embedding the nodes along a line and use this embedding in future optimization steps.  This approach has been proved to be effective for highly connected networks and expander graphs such as social networks~\cite{aydin2015distributed}. Our work in this paper focuses on clustering with a geographically defined network, in contrast to the focus upon social networks. Our contribution to this literature is the parallelization of the ``natural cuts'' algorithm \cite{DGRW12, DGRW11}, which has been shown to be effective in solving the balanced partitioning problem, particularly for geographical graphs.

\section{Algorithm}
\label{sec:algorithm}

We now present the GeoCUTS algorithm. The input is a set of locations and individual users who have issued queries there. The algorithm proceeds in two phases. In Phase 1, we build a graph from the given data by setting nodes to discretized locations and assigning edges between pairs of nodes that frequently share users. In Phase 2, we find a clustering of this graph by applying a geographic clustering algorithm that combines recently developed techniques. Both the graph-building and the graph-clustering algorithm are designed to be run massively in parallel.

\subsection{Phase 1: Graph building}\label{graphbuilding}

\subsubsection{Discrete locations.}
The first step in building our graph is location discretization: we round each location to the nearest gridpoint in a lattice, where the width of the lattice may be specified; a coarser lattice yields a faster but less precise algorithm. We have chosen to discretize to a grid, rather than for instance to the nearest city, as described in Ugander and Backstrom \cite{ugander2013balanced}, because a grid has a natural geometrical structure that we exploit in our algorithm. As a result, our method is applicable to location data which might include users in very rural areas or along transit corridors.

\subsubsection{Node weights.} We define a graph for which the nodes are the gridpoints above; in the following discussion, we shall identify each node with the corresponding range of user locations. The weight of a node is a measure of the number of user visits to that location; specifically, if a user visits node $A$ a total of $a$ times, then the user's influence upon $A$ is $\sqrt{a}$: $$\text{weight}(A) = \sum_{\text{user}\, u} \sqrt{\text{\# visits $u$ to $A$}}.$$ We use $\sqrt{a}$ as a slight normalization. Alternatively, it would be possible to increment the node's influence by any other normalization or $a$ itself. We explored normalization because some individual users may have much more location data available than others, which would bias the graph strongly towards these users to the exclusion of others. We compared the efficacy of the square root normalization against other possible heuristics in \S\ref{subsec:normalization}.

\subsubsection{Edge weights.} The edges in this graph correspond to the intensity of transit between nodes.  Specifically, if a user visits node $A$ a total of $a$ times, and node $B$ a total of $b$ times, then the user's influence upon edge $AB$ is $\sqrt{ab}$: $$\text{weight}(AB) = \sum_{\text{user}\, u} \sqrt{(\text{\# visits $u$ to $A$})\cdot (\text{\# visits $u$ to $B$})}.$$We use the geometric mean to increment edge weights since it is minimized when either endpoint is visited seldom, and is maximized when the endpoints are visited equally.

\subsubsection{Sparsity.} We retain only certain edges within the graph, specifically those for which the geographical distance between the two nodes is less than a given parameter. The reason for this trimming of edges is twofold: firstly, it greatly reduces the size of the graph, allowing the number of edges to be linear, instead of at worst quadratic, in the number of nodes, which greatly speeds up the algorithm. Secondly, edges between nodes at great geographical distance interfere with our goal of geographically local clusters.

\subsubsection{Coverage}
In some instances, the data may not cover the entire geographical region of interest.  For example, in clustering users within the USA, it may be uncommon to receive user locations from within Death Valley.  This means that certain possible nodes do not have any weight in the graph. In the interest of complete geographic coverage, GeoCUTS fills these gaps by creating new nodes with small weight.  In addition, for every two nodes that are geographically close, the algorithm creates an edge of low weight.  This enforces the geographic plausibility of output clusters even in areas of low data density.

\subsubsection{Normalization.}
As a final and critical step, we re-normalize the weights in the graph.  We consider the cases of re-normalizing the weight $x$ to $\sqrt{x}$ and to $\log(x)$.  As we discuss further in \S\ref{sec:results}, we observed best results, both visually and according to the metrics introduced in \S\ref{sec:stats}, after log-normalizing both node and edge weights, though the normalization of nodes was most critical. This is due to the long tail distribution of our data, with small geographical regions concentrating a (very) large amount of traffic. Without normalization, our algorithm would attempt to aggressively split large cities, for example, to achieve exact balance between clusters.

\subsection{Phase 2: Graph clustering}\label{graphclustering}

As a crucial step in many graph mining problems, graph clustering is an active research area and numerous algorithms have been developed for this problem. Since our objective is to minimize interference (that is, the interaction between clusters) while maintaining clusters with roughly similar size, we have chosen an algorithm that solves the {\em balanced partitioning problem}. While this problem is computationally hard even for small instances, a seek a distributed solution for large-scale graphs. We first formally define balanced partitioning and then present our distributed algorithm for this problem on geographic graphs.  

Our algorithm improves on the the natural cuts heuristic~\cite{DGRW12,DGRW11} by parallelizing two steps of the algorithm: the seed selection can be done in parallel using a Hilbert curve embedding (see \S\ref{hilbertcurveembedding}); the contraction around seed nodes can be done in parallel by using a distributed hash-table service~\cite{aydin2015distributed, KiverisLMRV14}.

\subsubsection{The balanced partitioning problem.}
Consider a graph $G(V, E)$ of $N$ vertices with edge weights $\ell : E \rightarrow R$, node weights $w: V \rightarrow R$. Let $w(S)$ for any subset $S$ of nodes be the total weight of nodes in $S$, i.e., $w(S) = \sum_{j \in S} w(j)$.  A partition of nodes of $G$ into $M$ parts $\{V_k: k\in [M]\}$ is said to be $\alpha$-balanced if and only if
$$\forall k \in [M],~(1-\alpha) \frac{w(V)}{M}  \leq w(V_k) \leq (1+\alpha)\frac{w(V)}{M}$$

In particular, a zero-balanced (or fully balanced) partition is one where all partitions have the same weight. The weight of the cut for this partitioning is the total sum of all edges whose endpoints fall in different regions $V_k$ and $V_l$: 
$$\sum_{k<l} ~ \sum_{(u, v) \in (V_k \times V_l) \cap E(G)} \ell(u,v).$$
In the balanced partitioning problem, the goal is to find an $\alpha$-balanced partitioning for which the cut size is minimized. In most cases, we are not given a specific $\alpha$ as input and must find a partitioning that is as balanced as possible, i.e., with minimum $\alpha$. For more statistical context on balanced partitioning, see \S\ref{sec:stats}.

Following the structure of algorithms based on natural cuts~\cite{DGRW12,DGRW11}, our distributed algorithm iterates on a contraction stage and generates the output by applying a post-processing contraction stage, in which we {\em merge} parts of the contracted graph to compute the final partitioning.
\begin{algorithm}
\SetAlgoLined
\SetKwData{Left}{left}\SetKwData{This}{this}\SetKwData{Up}{up}
\SetKwFunction{IdentifySeedSetMapRed}{\textbf{IdentifySeedSetMapRed}}
\SetKwFunction{NaturalCutsMapRed}{\textbf{NaturalCutsMapRed}}
\SetKwFunction{ConnectedComponentsMapRed}{ConnectedComponentsMapRed}
\SetKwFunction{GraphContractionMapRed}{GraphContractionMapRed}
\SetKwFunction{GraphAssembly}{\textbf{GraphAssembly}}
\SetKwFunction{OutputClusters}{OutputClusters}
\SetKwFunction{GetHilbertEmbedding}{GetHilbertEmbedding}
\SetKwFunction{SplitEqualParts}{SplitEqualParts}
\SetKwFunction{GetMiddleNodeOfEachParts}{GetMiddleNodeOfEachParts}
\SetKwFunction{ContractSubGraph}{ContractSubGraph}
\SetKwFunction{ExpandNeighborhood}{ExpandNeighborhood}
\SetKwFunction{ComputeSourceTargetCut}{ComputeSourceTargetCut}
\SetKwFunction{MatchBestPair}{MatchBestPair}
\SetKwFunction{ContractPair}{ContractPair}
\SetKwFunction{ClustersOf}{ClustersOf}
\SetKwInOut{Input}{Input}
\SetKwInOut{Output}{Output}

\KwIn{Undirected graph $G(V, E)$ with node and edge weights and desired cluster size $U$}
\KwOut{A partition of $G$ into clusters of size near $U$}

\BlankLine

\emph{\textbf{Graph Contraction:}}\

\Indp
    $S \leftarrow$ \IdentifySeedSetMapRed($G$, $U$)\;
   
    \hspace{15pt}  $Em\leftarrow $ \GetHilbertEmbedding($G$, $U$)\;
    
    \hspace{15pt}  $Parts\leftarrow $ \SplitEqualParts($U$, $Em$)\;
    
    \hspace{15pt}  $return$ \GetMiddleNodeOfEachParts($Parts$)\;
    
    $C \leftarrow$ \NaturalCutsMapRed($G$, $S$, $U$); \ //\,Cut edges
    
    \hspace{15pt}  $Q \leftarrow \frac{w(V)}{M} $\;
    
    \hspace{15pt}  $C(v) \leftarrow $ \ExpandNeighborhood($v$, $Q/10$)\;
    
    \hspace{15pt}  $s \leftarrow $ \ContractSubGraph($C(v)$)\;
    
    \hspace{15pt}  $G^\prime(v) \leftarrow $ \ExpandNeighborhood($s$, $Q$)\;
    
    \hspace{15pt}  $t \leftarrow $ \ContractSubGraph($G - G^\prime(v)$)\;
    
    \hspace{15pt}  $return$ \ComputeSourceTargetCut($G^\prime(v), s ,t$) \;

    $CC \leftarrow$ \ConnectedComponentsMapRed($G(V, E - C)$)\;

    $H \leftarrow$ \GraphContractionMapRed($G$, $CC$)\;
\Indm
\emph{\textbf{In Memory Merge:}}

\Indp
    $P  \leftarrow$ \GraphAssembly($H$)\;
    
    \hspace{10pt} $H^\prime \leftarrow H,$ $repeat:$
    
    \hspace{18pt}  $(u,v) \leftarrow $ \MatchBestPair($H^\prime$) \ // $w(u) + w(v) \leq U$
    
    \hspace{18pt}  $H^\prime \leftarrow $ \ContractPair($H^\prime, u, v$)\;
    
    \hspace{10pt} $return$ \ClustersOf($H^\prime$) \ // $H^\prime$ nodes are clusters
    
    \OutputClusters($P$)\;
\BlankLine
\BlankLine

\caption{Phase 2 of the GeoCUTS algorithm.}
\label{alg:graph-clustering}
\end{algorithm}

\subsubsection{Our algorithm.}\label{clustering-stages}
In this section we go into the details of the main stages of our distributed algorithm: 1) Contraction stage to reduce the graph size into a smaller one and 2) Merging stage that applies expensive heuristics to a small contracted in-memory graph.
\newline
{\bf \noindent Contraction stage.}
This stage consists of a sequence of four main steps: (i) identifying a seed set, (ii) finding a natural cut around each seed node, (iii) computing connected components of the graph after removing edges of the natural cuts, and (iv) finally contracting nodes in each connected component to one node and computing an updated smaller graph with new node weights and edge weights. The main contribution of our distributed implementation is that all of the natural cuts are computed in parallel, and the graph contraction based on these cuts also happens in a distributed manner.

\begin{enumerate}
\item{\bf Identify a seed set.} Here, the goal is to identify a set of $M$ seed nodes $S$ from which we compute natural cuts. We follow the following strategy for computing this set $S$: embed nodes of the graph into a line using the Hilbert curve (see \S\ref{hilbertcurveembedding}), and divide the line into $M$ pieces, each with almost the same number of nodes. Then, output a node close to the center of each piece of the Hilbert curve embedding.  This method ensures that the set of seed nodes are spread uniformly across different parts, and thus natural cuts cover different parts of the graph.

\item{\bf Natural cuts around seed nodes.}  After selecting the seed set, we compute a ``natural cut'' around each seed node in parallel.  Consider a seed node $v$, and let $Q~= (1+\alpha)\frac{w(V)}{M}$, i.e., $Q$ is the maximum weight of a cluster in an $\alpha$-balanced partitioning. The idea is first to compute a core $C(v)$ around node $v$ by performing a BFS around node $v$ until we cover $Q/10$ nodes (wher the constant $10$ is heuristic). We contract $C(v)$ to a node $s$. Then, we continue the BFS until the total size of the neighborhood reaches $Q$ and form a graph  $G'(v)$ around node $v$. We take the rest of the graph $G\backslash G'(v)$ and contract it to one node, denoted by $t$. Finally, we compute a minimum $(s,t)$-cut in this graph $G'(v)$. We call this $(s,t)$-cut a {\em natural cut} around seed node $v$. A desirable property of this cut is that it has less than $Q$ total weight on its nodes, and can be used as a building block for computing parts of a balanced partitioning.  We can compute all these natural cuts in parallel in a distributed manner by applying a MapReduce framework, uploading the graph in a distributed hash-table service, and accessing the neighborhood of nodes via a read-only service~\cite{aydin2015distributed, KiverisLMRV14}.

\item{\bf Distributed connected components.} As the next part of the contraction stage, we remove all edges of the graph that appear in at least one of the natural cuts computed around the seed nodes, and then compute connected components in the remaining graph (after removing those edges). We apply a distributed implementation of connected components that employs a distributed hash-table service that has been shown to be effective and scalable  in practice~\cite{KiverisLMRV14, RastogiMCS13}. 

\item{\bf Contraction of each connected component.}
After computing connected components, we can easily construct a contracted graph as follows: we put a node $u_i$ for each connected component $T_i$ with node weight $w(T_i)$, i.e.,  the sum of the weight of nodes in $T_i$. We also set the weight of the edge between two nodes $u_i$ and $u_j$ to the sum of edge weights between components $T_i$ and $T_j$.
\end{enumerate}

{\bf \noindent Merging stage.}
If the size of the contracted graph is large, we iterate on the contraction stage until the size of the graph is small enough to fit in memory.\footnote{For our data sets, we never needed to iterate on this stage, since after the first stage, the graph fits in memory.} When the contracted graph fits in memory, we can produce an output by applying any in-memory heuristic for balanced partitioning of graphs with node weights. The algorithm that we employ at this stage is similar to the greedy assembly algorithm proposed in~\cite{DGRW12,DGRW11}.
 
\subsubsection{Hilbert curve embedding}\label{hilbertcurveembedding} 
\begin{figure}[htb]
\centering
\includegraphics[scale=0.25]{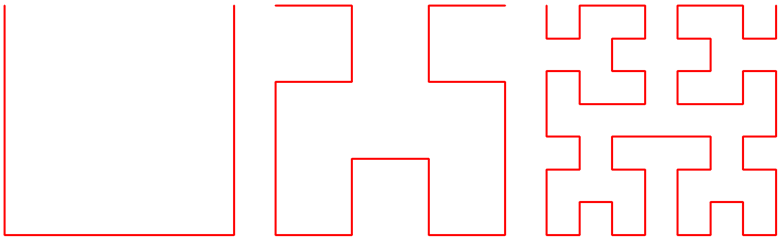}
\caption{Hilbert space-filling curves are constructed recursively up to any desired resolution.
\vspace{-3mm}
}
\label{fig:hilbert-curve}
\end{figure}

A Hilbert curve is a space-filling curve that has a fractal-like structure first described by German mathematician David Hilbert in~\cite{article:MJFS01:analysis-hilbert}.   Figure~\ref{fig:hilbert-curve} shows the first three steps of its construction. It can be recursively constructed up to any desired level to approximate a space by dividing it into cells. One of its most desirable properties is that close distances in 2D space also stay (mostly) close on the 1D line. This property can be used to find dense regions on a geographic graph by simply inspecting dense segments on the line. We use this property in our parallelization of the natural cuts seed selection (cf.~\S\ref{clustering-stages}). 

\section{Statistical evaluation}
\label{sec:stats}

In this section, we define a set of metrics to measure how well a given partitioning supports the purpose of experimentation. In particular, we introduce the Q-metric as an improvement over the oft-used graph-cut metric in the literature on experimentation with interference \cite{eckles2014design,gui2015network}.


\subsection{Quantifying interference: the Q-metric}
In many A/B tests, the units of experimentation can be considered independent: the outcome of one unit is affected only by whether it is assigned to treatment or control. In the potential outcomes framework~\cite{imbens2015causal}, we say that the Stable Unit Treatment Value Assumption (SUTVA)~holds, in which case the difference-in-means estimator (and other common estimators) is provably unbiased for the treatment effect estimand. 

However, in certain A/B tests, the outcome of one unit may be affected by the treatment status of units around them. In this case, we say there is interference: the Standard Unit Treatment Value Assumption (SUTVA) does not hold and our estimators are no longer guaranteed to be unbiased. In the geo-experiments case, the units of randomization are geographical regions; their outcomes are the aggregation of the user activity within each region. Because users travel from region to region, the outcome of one region does not depend only on whether that region is assigned to treatment or control, but also on the treatment status of all neighboring regions between which its users might travel.

The causal inference literature often represents interference by a graph on the experimental units, where an edge is drawn between two units likely to interfere with one another~\cite{eckles2014design, athey2018exact, zigler2018bipartite}. In this representation, two disconnected components of the graph (groups of regions with no users travelling from one to the other) do not affect each other's outcome. The edge weights of the graph are chosen to be representative of the interference structure and often the result of a domain-informed heuristic.

\begin{figure}[htb]
\includegraphics[scale=.3]{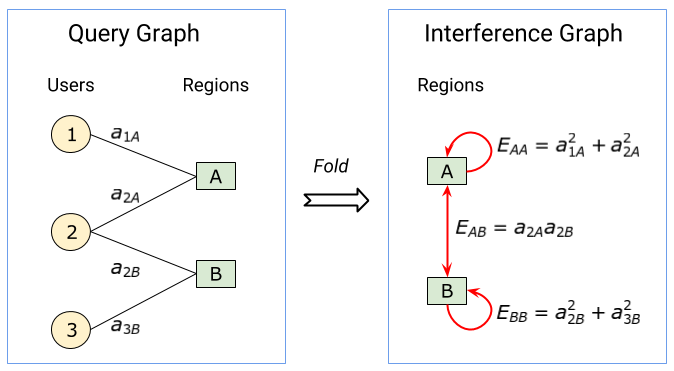}
\caption{Diagram of the bipartite user-region graph and the resulting ``folded'' interference graph between regions. The edge weights of the folded graph correspond to the unnormalized weights $q_{kk}$ (cf. Equation~\ref{eq:unweighted_edges}).
\vspace{-3mm}
}
\label{fig:bipartite}
\end{figure}

In our case, there is a clear underlying bipartite graph between users and regions (see Figure \ref{fig:bipartite}), from which we can build the interference graph between regions. Let $a_{ik}$ be the number of queries performed by user $i$ in region $k$. A natural weight to consider for a pair of geo-clusters $k$ and $k'$ is the folded edge $E_{kk'}$: 
\begin{equation}
\label{eq:unweighted_edges}
q_{kk'} = \sum_i a_{ik} a_{ik'} 
\end{equation}
Much like in the graph building step of the GeoCUTS algorithm (cf.~\S\ref{sec:algorithm}), we seek to normalize these edge weights to account for the large variance of information available across users and regions. Letting $a_{:k} = \sum_i a_{ik}$ and $a_{i:} = \sum_k a_{ik}$ be the region-aggregated and user-aggregated outcomes respectively, we consider the normalized folded edge:
$$Q_{kk'} = \sum_i \frac{a_{ik} a_{ik'}}{\sqrt{a_{:k}a_{:k'}} \sqrt{a_{i:} a_{i:}}}.$$

To measure the ``quality'' of each individual cluster or region, we can simply set $k=k'$, as is done in the following definition:
\begin{newdef}
We define the \emph{quality} of region $k$ by:
$$Q_k =\sum_i \frac{a_{ik}^2}{a_{:k}a_{i:}},$$
where $a_{i :}$ is the total number of queries issued by user $i$ across all clusters and $a_{: k}$ is the number of queries in region $k$ issued by all users. Let $M$ be the number of regions. We define the \emph{quality} or \emph{Q-metric} of the overall geo-clustering as the mean quality of each region: $$\bar{Q} = \frac{1}{M} \sum_k Q_k$$
\end{newdef}

The Q-metric is a natural extension of the graph-cut metric~\cite{gui2015network, eckles2014design} to the bipartite setting of users and regions. Much like a cut metric for interference,  $\bar{Q} = 1$ when regions are perfectly isolated and no user travels between two regions, and $\bar{Q} \sim \frac{1}{M}$ when each user participates equally in each region.

In constructing the Q-metric, we have considered regions to be units both of analysis and of randomization. Since our setting is bipartite, we could consider users to be units of analysis (but not randomization) as explored in~\cite{donner2004pitfalls,zigler2018bipartite}. We would then face the problem of modeling user response to varying levels of treatment over time. As a user travels from region to region, he or she will be exposed to various values of treatment. 

For example, assume that a user's response is proportional to their ``treatment dose'', i.e.~the ratio of queries made within treated regions over the total queries made. Let $Z_k \in \{0,1\}$ be the treatment status of region $k$; $Z_k = 1$ if treated and $Z_k = 0$ otherwise. Then, the treatment dose $d_i$ received by user $i$ and user $i$'s response $Y_i^t$ at time $t$ are given by:
\begin{equation}
\label{eq:dose}
d_i = \frac{\sum_k Z_k a_{ik}}{a_{i:}} \quad \text{and} \quad Y_i^t = Y_i^0 \left( 1 +  \beta d_i\right),
\end{equation}
where $Y_i^0$ is the response of user $i$ prior to the start of the experiment and $\beta \in \mathbb{R}$ is an arbitrary coefficient. 

In this dosed-response setting, we find evidence that the Q-metric is an appropriate measure of clustering quality for the purpose of experimentation with interference. Let $TE$ be the treatment effect estimand, defined as the difference between responses when every region is treated ($Z = \vec 1$) and when none is treated ($Z =\vec 0$):
$$TE = \frac{1}{M} \sum_i Y_i^t(Z = \vec 1) - \frac{1}{M} \sum_i Y_i^t(Z = \vec 0)$$
Though $Y_i^t$ could be modeled as a linear function of the number of queries $a_{i:}^t$ made by user $i$ at time $t$, we let $Y_i^t = a_{i:}^t$ to simplify the exposition of the following proposition, which serves to illustrate the relevance of the Q-metric.

\begin{proposition}\label{prop:linear}
  Under the linear-outcomes model for users given in Equation \ref{eq:dose}, the
  expectation of the difference-in-means estimator $\hat \tau$ for the treatment effect with respect to the assignment of regions to treatment and control is:
  $$\mathbb{E}_Z\left[ \hat \tau \right] = TE + \beta \left( \bar{Q} - 1\right)$$
\end{proposition}
\begin{proof}
Let $M_t$ be the number of regions assigned to treatment and $M_c$ the number of regions assigned to control, such that $M = M_t + M_c$. The difference-in-means estimator considers the difference of \emph{relative} responses of treated and control geo-regions:
$$\hat \tau = \frac{1}{M_t} \sum_k Z_k \frac{\sum_i a_{ik}^t}{a_{:k}^0}- \frac{1}{M_c} \sum_k (1 - Z_k) \frac{\sum_i a_{ik}^t}{a_{:k}^0}$$
Considering the user responses, linear in the treatment dose, and cancelling out the constant term, from Equation \ref{eq:dose}, the left-hand side (LHS) of the estimator becomes:
$$\hat \tau_{LHS} = \frac{1}{M_t} \sum_k \frac{Z_k}{a_{:k}^0} 
    \sum_i a_{ik}^0 \left( 1 + \beta \frac{\sum_{k'} a^0_{ik'} Z_{k'}}{a^0_{i:}} \right)$$
Taking the expectation with respect to the treatment assignment vector $\mathbf{Z}$ of the LHS estimator,
$$\mathbb{E}_Z \left[ \hat \tau_{LHS} \right] = \frac{1}{M} \sum_k \frac{1}{a_{:k}^0}
    \sum_i a_{ik}^0 \left(1 + \beta \frac{a_{ik}^0}{a^0_{i:}} + \frac{M_t}{M} \frac{\sum_{k' \neq k} a^0_{ik'}}{a^0_{i:}} \right)$$
Computing the difference with the expectation of the right-hand side (RHS) of the estimator,
$$\mathbb{E}_Z \left[ \hat \tau \right] = \frac{1}{M} \sum_k \frac{1}{a_{:k}^0}
    \sum_i a_{ik}^0 \beta \frac{a_{ik}^0}{a^0_{i:}} = \frac{1}{M} \sum_k Q_{kk} = \beta \bar Q$$
Since the treatment effect is given by $TE = \beta$, we recover the formula in Proposition~\ref{prop:linear}:
$$\mathbb{E}_Z\left[ \hat \tau \right] = TE + \beta \left( \bar{Q} - 1 \right)$$
\end{proof}

The Q-metric of the clustering is an appropriate measure for quantifying the quality of the geo-partitions in terms of interference: the higher $\bar Q$, the lower the bias of the difference-in-means estimator. In fact, when $\bar Q$ is at its maximum ($\bar Q = 1)$, the estimator is unbiased for the treatment effect. It is, to the best of our knowledge, the first heuristic of its kind for measuring cluster quality in a bipartite interference-graph setting. 

\subsection{Quantifying balance: the B-metric}
While the Q-metric is a good metric for measuring the interference present in an experiment, it cannot be the only yardstick by which we measure the overall quality of our experimental set-up. If it were, we could place the majority of users in one very large cluster, letting other clusters be sparsely populated (e.g. the contiguous United States vs. Hawaii and Alaska). While this would achieve a high Q-metric score, the variance of our estimator would be very large since our estimate would be strongly dependent on the treatment status of the large cluster.

As an illustrative example, suppose the response rate of individual users is given by Eq.~\ref{eq:dose} and that a single cluster $k$ concentrates far more queries than any other clusters: 
\begin{equation*}
\sum_i a_{ik} >> \sum_i \sum_{k' \neq k} a_{ik'}
\end{equation*}
If cluster $k$ is placed in treatment, the treatment effect estimate will be approximately $\beta \cdot \sum_i a_{ik}$, otherwise it will be approximately $\beta \sum_i \sum_{k' \neq k} a_{ik'}$. As a result, the variance of our estimator is:
\begin{align*}
    \text{var}_\Z[\hat \tau] \approx \frac{\beta^2}{8} \left( 2 a_{:k} - a_{::} \right)^2 + \frac{\beta^2}{8} \left(a_{::} - 2a_{:k} \right)^2 \approx \frac{\beta^2}{4} a_{:k}^2
\end{align*}
where $a_{:k} = \sum_i a_{ik}$ and $a_{::} = \sum_i \sum_k a_{ik}$. We have assumed that $a_{:k} \approx a_{::}$ in the above comparisons, i.e. there exists one large cluster that concentrates the vast majority of user queries.

Despite the constant treatment effect parameter $\beta$ on each individual user, the variance of our estimator is large because one cluster is responsible for determining the treatment dose received by an overwhelming majority of users. To avoid such a scenario, we introduce the following balance metric.

\begin{newdef}
Let $w_k$ be the f-normalized weight of geo-region $k$:
$$ w_k = \frac{\sum_i f(a_{ik})}{\sum_i \sum_{k' \neq k} f(a_{ik'})}$$
where $f$ is a normalization function of our choosing. We define the \emph{B-metric} of the clustering as the quantity 
$$\left( \|w\|^2 - \frac{1}{M} \right)$$
where $M$ is the total number of geo-regions.
\end{newdef}

As in the graph-building phase described in \S\ref{sec:algorithm}, we explored both $\sqrt{\cdot}$ and $\log(\cdot)$ and the identify function as normalizations (cf.~Table~\ref{tab:qandbmetrics}). As expected, the B-metric is equal to 0 if the clustering is perfectly balanced and is otherwise positive, with greater size indicating greater imbalance.

\subsection{Effective number of clusters}
Our proposed balanced partitioning algorithm (cf. ~\S\ref{sec:algorithm}) optimizes both Q- and B-metrics in an effort to mitigate interference while enforcing balance. In achieving this trade-off, the number of clusters must be specified. As a rule of thumb, the larger the number of clusters, the more difficult it will be to attain a high Q-metric. In the case of a complete interference graph for example, the highest achievable Q-metric is $\frac{1}{M}$, the inverse of the number of clusters specified by the user.

While having few clusters may achieve a high Q-metric score without necessarily impacting the B-metric, having few experimental units has other undesirable statistical properties: high variance of estimators, low coverage of confidence intervals, covariate imbalance, etc. Ultimately, the ideal number of clusters will be experiment-dependent. We suggest, in practice, evaluating covariate imbalance, running AA tests, and simulating possible user responses in order to determine experimental power under various numbers of clusters. In our experiments (cf. Table~\ref{tab:qmetric}), we have used, as a baseline, a fixed number of clusters based on that for established hand-designed geo-regions.

\section{Empirical Results}
\label{sec:results}
In this section we evaluate our algorithm and compare it against alternative algorithms and baselines.

\subsection{Dataset}
\label{subsec:data}
\begin{figure}[htb]
\centering
\includegraphics[scale=0.18]{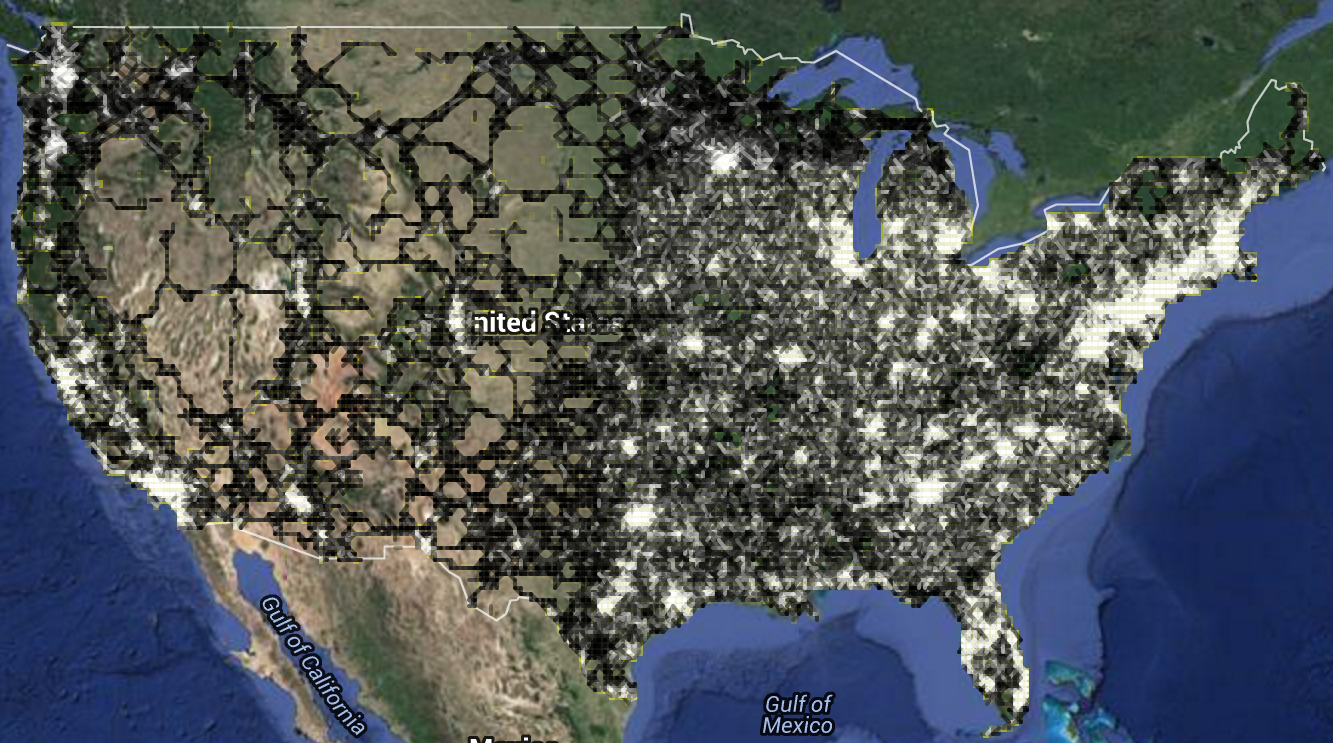}
\caption{The graph built by the GeoCUTS algorithm, with nodes shown on corresponding locations of the US. White represents large edge weights (high traffic areas), while black represents low edge weights. Larger edge weights often do not match larger vertex weights, showing the difference between GeoCUTS and an algorithm simply measuring population density. Gaps in the colored regions represent locations for which no search data is available. For example, in regions such as deserts, Search queries come disproportionately from narrow strips corresponding to major roads.
\vspace{-7mm}
}
\label{fig:us-graph}
\end{figure}

We seek in our clustering to minimize the interference introduced by movement of users; hence, identifying movement trends is crucial. A natural choice for reconstructing movement of users is the (approximate) location of Google Search queries. We use a massive dataset consisting of 1 percent of all Google Search data over a period of 28 days. Our data from Search queries consists of an anonymized set of browser cookies, each associated with a number of locations at which that cookie has issued search queries. The locations are approximate (both of necessity and to preserve anonymity) and specified as a bounding box that covers the location at which a query was issued. The size of the bounding boxes is not uniform but all of them are large enough to contain locations of queries issued by a large number of distinct cookies. Larger boxes tend to occur in rural areas, where geo-location is less accurate and where the smaller number of queries also makes it necessary for bounding boxes to be larger in order to ensure anonymization. Due to the scale of our datasets, while we have only a rough position for each individual query, and potentially only a few queries per cookie, the aggregate movement patterns are quite accurate.

\begin{figure}[htb]
\centering
\includegraphics[scale=0.09]{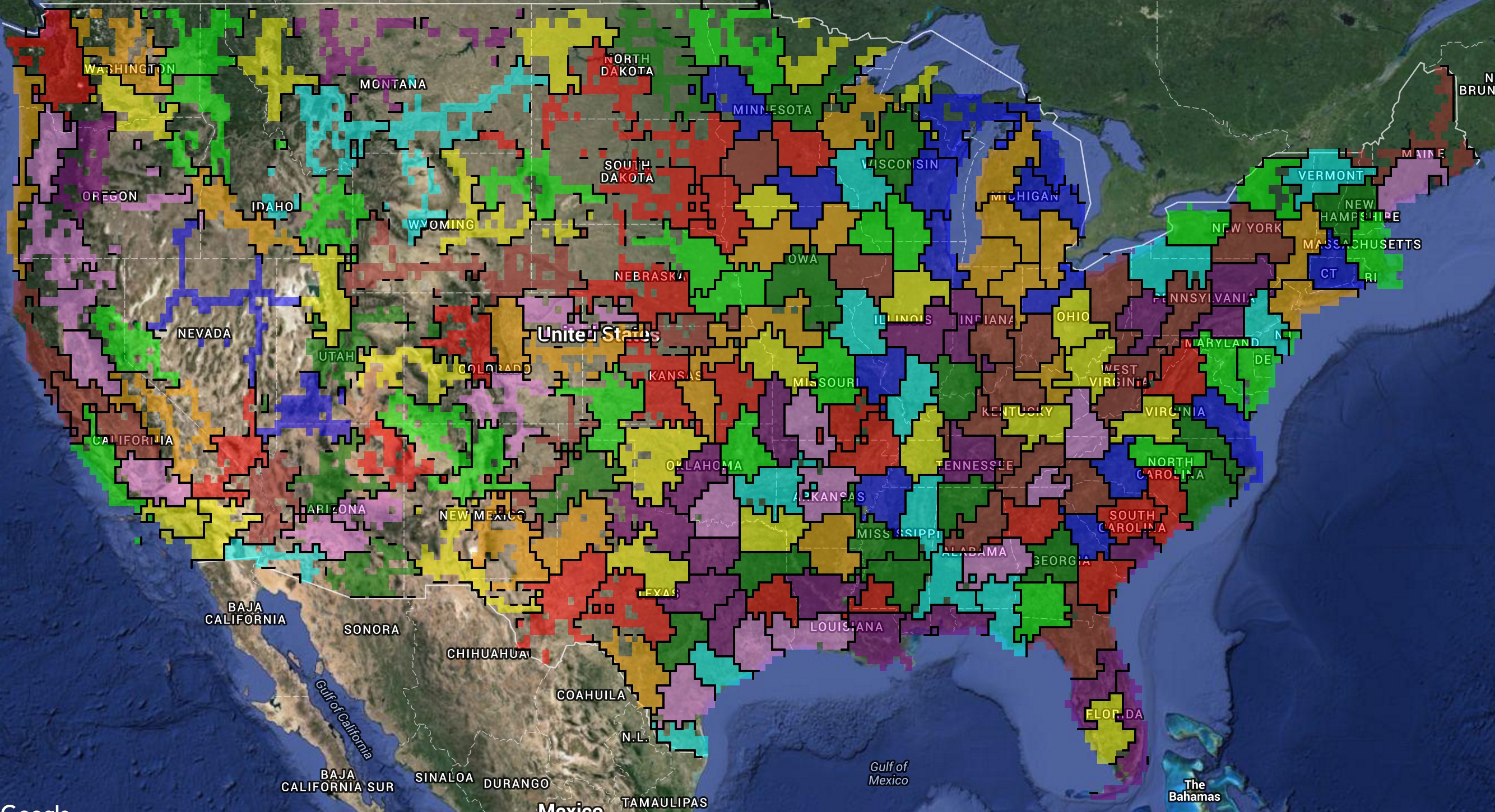}
\caption{The GeoCUTS algorithm applied to user queries from the United States. The algorithm automatically identifies metropolitan areas, correctly predicting, for example, that the Bay Area includes San Francisco, Berkeley, and Palo Alto, but not Sacramento.
\vspace{-3mm}
} \label{fig:us-clusters}
\end{figure}

To build a graph, we form a grid on the geographical area we wish to partition. Each grid cell is a node in our graph, with the edge between two nodes weighted based on the number of cookies that issue queries in both corresponding grid cells. Thus, for each query we need only identify the cell it is issued from; for this purpose, we assume that each query is issued at the center of its bounding box. To ensure that inaccuracies in estimating positions do not negatively impact our algorithm, grid cells must be large relative to the typical sizes of bounding boxes. We will take this assumption into account when we discuss the granularity of grid cells.

\begin{figure}[htb]
\centering
\includegraphics[scale=0.09]{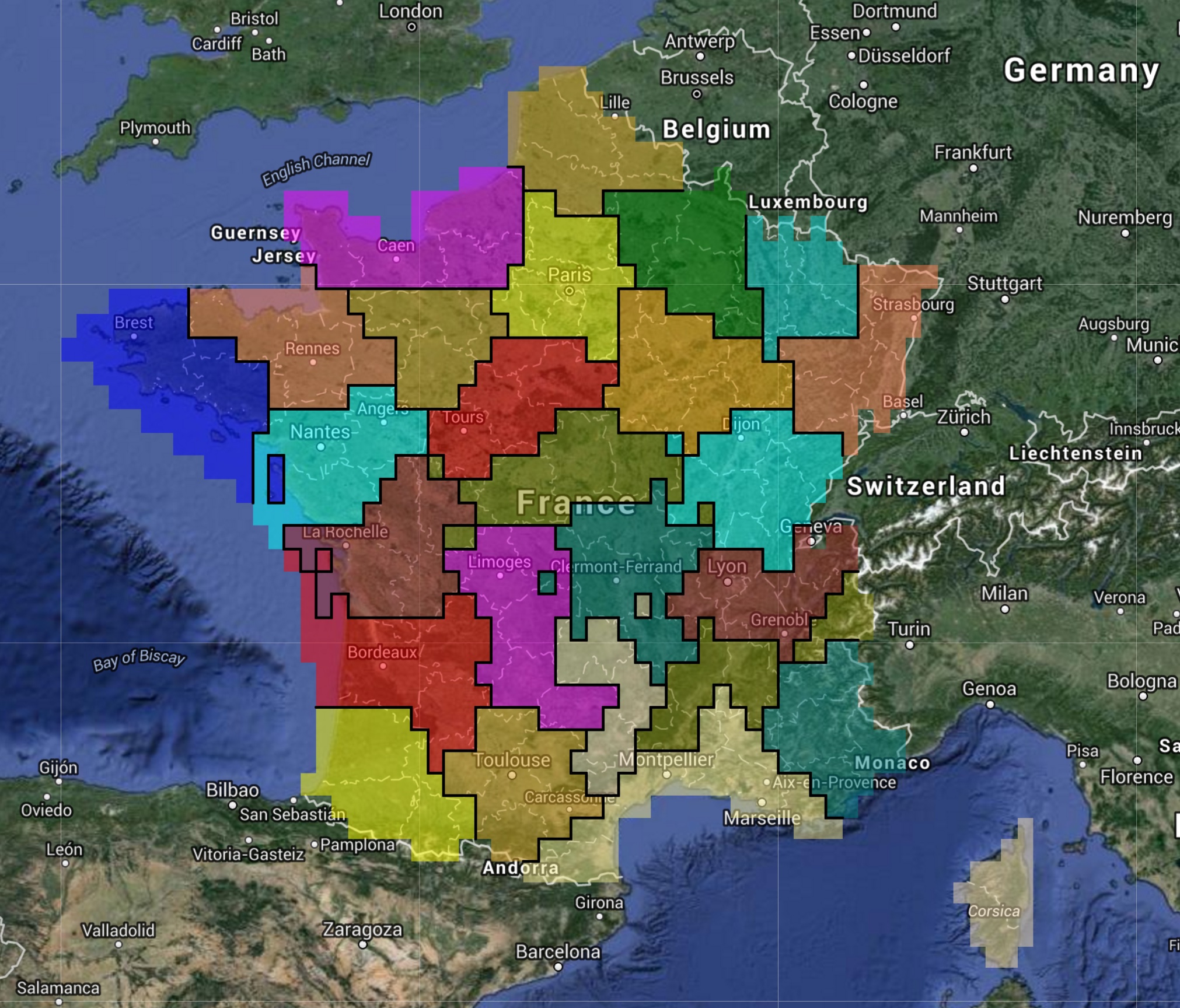}
\caption{The GeoCUTS algorithm applied to user queries from France. It correctly identifies metropolitan areas such as Paris, Bordeaux, and Lyon, and regions such as Alsace and Normandy.
\vspace{-1mm}
} \label{fig:fr-clusters}
\end{figure}

In the rest of this section, we evaluate the performance of the GeoCUTS algorithm on different geographic regions, for each of which a separate graph was built and clustered. Figures~\ref{fig:us-clusters} and~\ref{fig:fr-clusters} show the clusters generated for data from the United States and France, respectively, and Figure~\ref{fig:us-graph} shows the GeoCUTS graph before clustering. Note that prior hand-designed geo-regions have largely focused on the United States and, unlike GeoCUTS, cannot be generalized to other regions without extensive additional labor.

\subsection{Mobility}
\label{subsec:mobility}

Stationary cookies (cookies that do not move) are not interesting for our problem. In an extreme scenario where all cookies are stationary and issue queries from one location only, any arbitrary clustering algorithm performs perfectly well in terms of interference. Hence for our first dataset we consider only \emph{highly mobile} cookies: cookies that issue queries in at least two different cells of the grid.

The typical time-scale of experiments ranges from a few weeks to a few months, just long enough for a response equilibrium to be reached and for metrics to stabilize. While migrations may not affect the result of a one-day experiment, interference effects become more pronounced as time passes. Furthermore, some cookies are churned rather shortly after they are created; thus, multiple cookies may represent the same user over the period of our analysis. Hence, cookies with low query frequency under-represent the true movement. Query bounding boxes are in fact samples from the actual movement path, and a small number of samples is not enough to reconstruct the path. Hence, for our second dataset we consider {\em highly active} cookies only: cookies which issue a query in more than 10 out of the 28 days.
\begin{table*}[htb]
\centering
\begin{tabular}{cccccccc}
 (a) & & \multicolumn{2}{c}{\bf GeoCUTS}   & \multicolumn{2}{c}{\bf DMA} &
  \multicolumn{2}{c}{\bf Grid}   \\
 & & Avg & Query-w. avg& Avg & Query-w. ag& Avg & Query-w. avg\\
\hline
US&Highly Active & $87\%$ & $92\%$ & $88\%$ & $92\%$ & $84\%$ & $91\%$ \\
&Highly Mobile& $79\%$ & $85\%$ & $80\%$ & $85\%$ & $76\%$ & $81\%$   \\
France&Highly Active& $84\%$ & $89\%$ &-&-&$83\%$  &$86\%$ \\
&Highly Mobile & $74\%$ & $79\%$ & - & - & $75\%$ & $77\%$\\
  & \\
\end{tabular}
\begin{tabular}{ccccccccccc}
 (b) & & \multicolumn{3}{c}{\bf GeoCUTS}   & \multicolumn{3}{c}{\bf DMA} &
  \multicolumn{3}{c}{\bf Grid}   \\
 & & $\geq 0.75$ & $\geq 0.8$ & $\geq 0.85$ & $\geq 0.75$ &  $\geq 0.8$ & $\geq
  0.85$ & $\geq 0.75$ &  $\geq 0.8$ & $\geq 0.85$\\
\hline
US&Highly Active & $100\%$ & $100\%$ & $97\%$ & $100\%$ & $100\%$ &
  $98\%$ &$100\%$ &  $99\%$ & $94\%$\\
&Highly Mobile& $96\%$ & $86\%$ & $52\%$ & $95\%$ & $81\%$ & $49\%$
  & $94\%$ & $60\%$ & $11\%$  \\
France&Highly Active & $100\%$ & $97\%$ & $80\%$ &-&-&- &$100\%$ &
  $89\%$ & $64\%$\\
&Highly Mobile&$78\%$ & $42\%$ & $11\%$ & -&-&-& $74\%$ & $24\%$ & $5\%$\\
\end{tabular}
  \caption{(a) Average and query-weighted average (Query-w.~avg) of Q-metric,
  (b) Percentage of queries from clusters with a Q-metric of at least $x\%$.
  $\sim 200$ clusters were used for the US and $\sim 50$ for
  France. For both highly active and highly mobile graphs, GeoCUTS performs comparably to DMAs and outperforms the baseline grid clustering.
\vspace{-3mm}
}
\label{tab:qmetric}
\end{table*}
Some highly active users may still have a limited movement and issue queries from the same geographical area. On the other hand, while some highly mobile users may issue fewer queries, they tend to move over a wider range. Therefore we expect interference to be higher for highly mobile users, which is validated in our experiments.

For each of the United States and France, we collect two datasets  - one for highly mobile and one for highly active cookies - and form a graph for each dataset. Both graphs have the same number of nodes (e.g.~about 11,000 for the US at grid size $0.25$ degrees). Unless otherwise specified, node and edge weights are log-normalized. In \S\ref{subsec:normalization}, we compare various normalization methods.



\subsection{Comparison against other clusterings}
\label{subsec:q-metric-cutsize}
In this section, we show that GeoCUTS regions are comparable to hand-designed geo-regions, while requiring no manual effort and extending naturally to different regions of the world and granularities. We compare GeoCUTS against the most popular set of hand-designed geo-regions in the US, Nielsen's DMAs (Designated Market Areas) \textregistered\,, which were created by Nielsen to correspond to television audiences. DMAs have the advantage of being well-established as a means of subdividing a user population. However, they are restricted to the US without a direct international equivalent, and their granularity is fixed, with approximately 200 regions.

We previously defined the Q-metric to quantify the interference within a clustering. In Table \ref{tab:qmetric}, we compare the Q-metrics of the output of GeoCUTS against DMAs (within the US) and also against a baseline automatic clustering consisting of a simple grid subdivision (within both the US and France).

The average and query-weighted average of the Q-metric for each clustering algorithm are shown in Table~\ref{tab:qmetric}(a). For each metric, GeoCUTS beats the baseline grid clustering. Where applicable, GeoCUTS and DMAs perform similarly well. It is important to note that in every evaluation, we compared only clusterings with similar numbers of clusters. Thus, in constructing the grid baseline, we picked the coarseness of the grid so that the number of regions in the grid approximated the number of clusters formed by GeoCUTS.  We also used $\sim 200$ clusters for GeoCUTS in the US, in order to provide an effective comparison with DMAs.

The fraction of clusters and queries for different lower bounds of Q-metric are shown in Table~\ref{tab:qmetric}(b). For example $80\%$ of queries in the highly active set are issued from minimum-cut clusters with a Q-metric of at least $0.8$. For the grid, $62\%$ of queries are issued from clusters with a Q-metric of at least $0.8$. As already noted, highly mobile graphs are more challenging to partition compared to highly active graphs. The data indicates that the gap between our algorithm and baseline is larger for highly mobile graphs.

While the Q-metric quantifies the interference, we must also compare the clustering algorithms in terms of balance. An algorithm that produces highly unbalanced clusters may outperform other alternatives if only the Q-metric is considered. For example, if we partitioned the US into 200 clusters where 199 of them were in Alaska and one cluster covered the rest of the country, we would obtain an almost perfect Q-metric as relatively few users would cross between clusters. However, such a clustering would not be useful for our applications. We compare B-metrics in Table~\ref{tab:bmetric}. The results indicate that, in terms of balance, GeoCUTS performs equally to the alternatives for highly active graphs, and slightly better for the highly mobile graph. In summary, while we perform better in terms of interference, we do not compromise balance.
\begin{table}[htb]
\begin{tabular}{cccc}
& {GeoCUTS}  & {DMA} & {Grid} \\
\hline
US Highly Active& 1.5 & 1.5 & 1.5 \\
US Highly Mobile& 1.8 & 1.7 & 1.3 \\
France Highly Active & 11.1 & - & 11.5\\
\end{tabular}
\caption{B-metrics across clusterings, reported with a multiplicative constant of $100$. We see that GeoCUTS performs comparably to other clusterings for highly active users, and somewhat better for highly mobile users.
\vspace{-3mm}
}
\label{tab:bmetric}
\end{table}

\begin{table}[htb]
\begin{tabular}{cccccc}
& GeoCUTS & DMA & Grid & LE & Hilbert \\ \hline
 HA  &  \textbf{4\%}  &  7\% &   15\%  & 4\% & 7\%  \\
 HM &  \textbf{4\%} &  7\%  &   14\% & 4\%  &  7\% \\
\end{tabular}
\caption{Cut size comparison against different clustering algorithms for highly active (HA) and highly mobile (HM) users within the US. ``Grid'' denotes the grid partition, ``LE'' denotes the Linear Embedding algorithm \cite{aydin2015distributed}, and ``Hilbert'' denotes partitions along a Hilbert curve \cite{article:MJFS01:analysis-hilbert}. We see that GeoCUTS and Linear Embedding give the best cut size.
\vspace{-3mm}
}
  \label{tab:cut-size}
\end{table}
\begin{table*}[htb]
\centering
\begin{tabular}{ccccccc}
 & \multicolumn{3}{c}{\bf GeoCUTS}   & \multicolumn{3}{c}{\bf Grid}   \\
 & $\geq 0.75$ & $\geq 0.8$ & $\geq 0.85$ & $\geq 0.75$ &  $\geq 0.8$ & $\geq
  0.85$ \\
\hline
{\bf $\sim25$ clusters}  & & & & & &\\
Highly Active & $100\%$ & $100\%$ & $94\%$&$100\%$ & $99\%$ & $69\%$\\
Highly Mobile& $94\%$ & $56\%$ & $10\%$ & $83\%$ & $30\%$ & $0\%$   \\ \hline
{\bf $\sim50$ clusters}  & & & & & &\\ 
Highly Active & $100\%$ & $97\%$ & $80\%$  &$100\%$ & $89\%$ & $64\%$\\
Highly Mobile& $78\%$ & $42\%$ & $11\%$ & $74\%$ & $24\%$ & $5\%$  \\
\end{tabular}
\caption{Percentage of queries from clusters with a Q-metric $\ge x\%$ for
  different numbers of clusters in France.
\vspace{-5mm}
}\label{tab:fr_no_clusters_qmetric}
\end{table*}

Finally, we compare GeoCUTS to other clusterings with respect to cut size (see Table~\ref{tab:cut-size}) on the US log-normalized graph with grid size $0.25$. It is clear that the GeoCUTS algorithm produces better cut sizes compared to DMA and grid partitioning. We also compare against Linear Embedding~\cite{aydin2015distributed} and against partitions generated along a Hilbert curve \cite{article:MJFS01:analysis-hilbert}, which the GeoCUTS algorithm similarly outperforms.

\subsection{Tuning the algorithm}
\label{subsec:normalization}
In this set of experiments, we consider how various design choices in the GeoCUTS algorithm affect performance. First, we consider different types of normalization during the graph-building phase (see Table \ref{tab:qandbmetrics}(a)). Specifically, we build graphs over US queries using logarithmic normalization of both vertices and edges, square root normalization, and also no normalization step at all. As expected, a stronger normalization is associated with better Q-metrics but worse B-metrics, demonstrating that normalization may be seen as mediating the trade-off between diminished interference and increased balance.
\begin{table}[htb]
\begin{tabular}{cccc}
  Normalization & $\log(\cdot)$ & $\sqrt{\cdot}$ & None \\ \hline
 Q-metric, Highly Active & \textbf{0.921}  &  0.881 & 0.840 \\
 Q-metric, Highly Mobile &  \textbf{0.854} & 0.807 & 0.765 \\ 
 B-metric, Highly Active & 1.65 & 0.47 &  \textbf{0.06}  \\
 B-metric, Highly Mobile & 1.82 &  0.53  &  \textbf{0.11} \\ 
  & \\
 Coarseness & 0.1 & 0.25 & 0.5 \\ \hline
 Q-metric, Highly Active & 0.916 & \textbf{0.921} & 0.858 \\
 Q-metric, Highly Mobile & 0.847 & \textbf{0.854} & 0.781 \\ 
 B-metric, Highly Active & 1.57 & 1.65 &  \textbf{1.15}  \\
 B-metric, Highly Mobile & 1.75 &  1.82  &  \textbf{1.32} \\ 
\end{tabular}
\caption{Comparison of weighted average Q-metrics and B-metrics for GeoCUTS applied to US query data across (a) varying normalizations, (b) varying coarsenesses of location discretization. The B-metrics are reported with a multiplicative factor of $100$.
\vspace{-5mm}
}
  \label{tab:qandbmetrics}
\end{table}

Next, we compare the performance of GeoCUTS across varying coarsenesses of grid cells, considering log-normalized graphs in which locations (latitude and longitude) are discretized to 0.1, 0.25, and 0.5 degrees (see Table \ref{tab:qandbmetrics}(b)).  It is worth noting that all of these sizes are considerably larger than the side length of a typical bounding box for location data. The coarsest discretization of 0.5 performs the worst in Q-metric and best in B-metric, as coarser discretization enforces balance but reduces the ability to decrease interference. In all other experiments, we have made a trade-off between the two metrics by rounding to the nearest 0.25 degree.

Finally, we consider the effect of varying the number of clusters (see Table \ref{tab:fr_no_clusters_qmetric}). As we noted in \S\ref{sec:stats}, decreasing the number of clusters increases the Q-metric. We see that GeoCUTS significantly outperforms the baseline grid regardless of the number of clusters.

\section{Conclusion}
\label{sec:conclusion}
We have presented an algorithm, GeoCUTS, for clustering user queries into geographical regions. These regions can be used to run cluster-based randomized experiments for measuring users' response under treatment. Clustering users geographically offers two major advantages: 1) assigning identical treatments to different browser cookies of the same user, and 2) mitigating the interference effects of interactions between users assigned to different treatments. Unlike existing systems, GeoCUTS can be run in any region of the world and for any number of clusters.  Alongside our clustering algorithm, we have introduced quality metrics for interference and balance of a given clustering for the purpose of running cluster-based randomized experiments. We evaluate GeoCUTS on these metrics, showing that it outperforms balanced partitioning baselines, and performs comparably to the state-of-the-art in hand-designed clustering.

\section*{Acknowledgments}
The authors would like to thank Kay Brodersen and Hal Varian for helpful advice. We also thank the Google New York graph mining team and especially Aaron Archer, Hossein Bateni, and Silvio Lattanzi. The research was carried out at Google, Inc. D.R.~was additionally supported by NSF Grant No.~1122374.
\bibliographystyle{abbrv}
\bibliography{references}
\end{document}